\documentclass[useAMS,usenatbib]{mn2e}
\usepackage{amsmath}
\usepackage{amsfonts}
\usepackage{amssymb}
\usepackage{amstext}
\usepackage{float}
\usepackage{graphicx}
\usepackage{subfigure}
\usepackage[abs]{overpic} 
\usepackage{rotating} 

\newcommand       \apj          {ApJ}
\newcommand       \apjl         {ApJL}
\newcommand       \aap          {A\&A}
\newcommand       \nat          {Nature}
\newcommand       \mnras        {MNRAS}
\def\simlt{\mathrel{\hbox{\rlap{\hbox{\lower4pt\hbox{$\sim$}}}\hbox{$<$}}}}
\def\simgt{\mathrel{\hbox{\rlap{\hbox{\lower4pt\hbox{$\sim$}}}\hbox{$>$}}}}
\def\lesssim{\mathrel{\hbox{\rlap{\hbox{\lower4pt\hbox{$\sim$}}}\hbox{$<$}}}}
\def\gtrsim{\mathrel{\hbox{\rlap{\hbox{\lower4pt\hbox{$\sim$}}}\hbox{$>$}}}}

\def\kms{{\rm\,km \, s^{-1}}}
\def\ergss{{\rm\,ergs\, s^{-1}}}
\def\msun{{\rm\,M_\odot}}

\title[The Lightcurves and Spectra of AIC]{Nickel-Rich Outflows Produced by the Accretion-Induced Collapse of White Dwarfs:  Lightcurves and Spectra}
\author[Darbha et al.]{S. Darbha$^{1}$\thanks{E-mail:siva.darbha@berkeley.edu}, B.~D. Metzger$^{2}$, E. Quataert$^{1}$, D. Kasen$^{3}$, P. Nugent$^{4}$, and R. Thomas$^{4}$ \\
$^{1}$ Astronomy Department and Theoretical Astrophysics Center, University of California, Berkeley, 601 Campbell Hall, Berkeley CA, 94720 \\ $^{2}$ Einstein Fellow; Department of Astrophysical Sciences, Peyton Hall, Princeton University, Princeton, NJ 08544, USA  \\ $^{3}$  Hubble Fellow; University of California, Santa Cruz, CA 95064, USA \\ $^{4}$  Computational Cosmology Center, Lawrence Berkeley National Laboratory, 1 Cyclotron Road MS50B-4206, Berkeley, CA, 94720}
\begin{document}
\date{Accepted . Received ; in original form }
\pagerange{\pageref{firstpage}--\pageref{lastpage}} \pubyear{2010}
\maketitle
\label{firstpage}

\begin{abstract}
The accretion-induced collapse (AIC) of a white dwarf  to form a neutron star can leave behind a rotationally supported disk with mass of up to $\sim 0.1 \msun$.  The disk is initially composed of free nucleons but as it accretes and spreads to larger radii, the free nucleons recombine to form helium, releasing sufficient energy to unbind the remaining disk.   Most of the ejected mass fuses to form $^{56}$Ni and other iron group elements.  We present spherically symmetric radiative transfer calculations of the transient powered by the radioactive heating of this ejecta.  We estimate the ejecta composition using nucleosynthesis calculations in the literature and explore the sensitivity of our results to uncertainties in the ejecta kinematics.  For an ejecta mass of $10^{-2} \msun$ ($3 \times 10^{-3} \msun$), the lightcurve peaks after $\lesssim 1$ day with a peak bolometric luminosity $\simeq 2 \times 10^{41} \ergss$ ($\simeq 5 \times 10^{40} \ergss$); the decay time is $\simeq 4 \, (2)$ days.  Overall, the spectra redden with time reaching $U-V  \simeq 4$ after $\simeq 1$ day; the optical colors ($B-V$) are, however, somewhat blue.  Near the peak in the lightcurve, the spectra are dominated by Doppler broadened Nickel features, with no distinct spectral lines present.  At $\sim 3-5$ days, strong Calcium lines are present in the infrared, although the Calcium mass fraction is only $\sim  10^{-4.5}$.  If rotationally supported disks are a common byproduct of AIC, current and upcoming transient surveys such as the Palomar Transient Factory should detect a few AIC per year for an AIC rate of $\sim 10^{-2}$ of the Type Ia rate.  We discuss ways of distinguishing AIC from other rapid, faint transients, including .Ia's and the ejecta from binary neutron star mergers.  

\end{abstract}


\section{Introduction}
\voffset=-1cm
\vspace{0.2 cm}
\label{sec:int}
\label{sec:intro}

As an accreting white dwarf (WD) approaches the Chandrasekhar mass, it suffers one of two fates:  (1) collapse to form a neutron star (NS) or (2) explosion as a Type Ia supernova (SN).  Which of these end states is realized in different astrophysical scenarios is not fully understood (see ~\citealt{Canal&Gutierrez97} for a review).  The composition of the WD is, however, a critical factor:  the relatively low densities required for electron captures on Ne makes accretion-induced collapse (AIC) of a WD to a neutron star possible, and thus O-Ne WDs, rather than C-O WDs, preferentially undergo AIC (\citealt{Nomoto84}; \citealt{Miyaji+80}; \citealt{Canal+90}; \citealt{Canal+92}; \citealt{Gutierrez+96, Gutierrez+05}; \citealt{Wanajo+09}).  AIC may result both from accretion following Roche-lobe overflow from a non-degenerate companion (the ``single degenerate'' AIC channel; e.g.~\citealt{Nomoto+79}; \citealt{Nomoto&Kondo91}), as well following the merger of two WDs in a binary system (e.g.~\citealt{Yoon+07}).

There is no direct observational evidence for AIC and so its rate remains uncertain.   Numerical simulations of AIC find that it is accompanied by a weak explosion ($\simeq 3 \times 10^{49}$ ergs) that synthesizes only a very small amount of Ni ($\lesssim 10^{-3} \msun$ in current models) \citep{WB,Dess06,Ott10}, unless there is a strong, large-scale magnetic field in the progenitor white dwarf \citep{Dess07,Metzger+08}.  The neutrino-driven wind from the proto-neutron star does produce $\sim 3 \times10^{-3} \msun$ of neutron-rich elements (specifically, matter with an electron fraction $\lesssim 0.4$; \citealt{Dess06}).  The low total abundance of such elements in the galaxy implies an AIC rate of $\lesssim 1 \%$ of the Type Ia rate, if the calculations of \citet{Dess06} are representative of AIC in general.   On empirical grounds, \citet{QianWass07} have suggested that AIC may in fact be the origin of the third peak $r$-process elements.

Because a WD accretes angular momentum as well as mass, it is likely to be rapidly rotating as it approaches $M_{ch}$.  Evidence in favor of this expectation is provided by the recent discovery that the X-ray pulsator RX J0648.0-4418 is an accreting WD with a mass $\gtrsim 1.2M_{\sun}$ and a rotational period $P \sim 13$ seconds \citep{m09}.  Rapid rotation of the WD prior to collapse would make AIC a strong source of gravitational wave emission \citep{Ott10}.   An electromagnetic counterpart to such a detection would significantly increase its scientific yield.  More generally, the detection of AIC would provide important new constraints on binary WD evolution and the origin of Type Ia SNe.  New optical transient surveys operating with high cadence and wide fields of view (e.g., Palomar Transient Factory [PTF; \citealt{Rau+09,Law+09}] and the Pan-STARRs-1 Medium Deep Survey) are sensitive to much fainter transients than previous surveys -- AIC may thus finally be detectable.   In order to increase the probability of such a detection, it is critical to have diagnostics that can isolate AIC among the deluge of transients that are being discovered.

In this paper, we present Monte-Carlo calculations of the optical-infrared lightcurves and spectra of transients produced by AIC.   Our radiative transfer calculations are motivated by the model presented in \citet{MPQ09-Nickel}.   They argued that the observational signature of AIC can, in some circumstances, be dominated by the ejecta produced by an accretion disk around the newly formed neutron star, rather than by the weak SN that accompanies the AIC itself.  To motivate our radiative transfer calculations, we first summarize \citet{MPQ09-Nickel}'s calculation of the disk's evolution and nucleosynthesis (\S \ref{sec:summary}).  We then describe our radiative transfer calculations (\S \ref{sec:fidmodcode}) and the composition and structure of the disk ejecta (\S \ref{sec:fidmodics}).   Our main results are presented in \S \ref{sec:results}.   We conclude by discussing the key properties of AIC, and comparing them to other predicted faint, rapid transients (\S \ref{sec:discussion}).

\vspace{-0.4cm}
\subsection{Summary of the Model}

\label{sec:summary}

Depending on the rotation profile of the WD prior to collapse, and in particular the degree of differential rotation (\citealt{Saio&Nomoto04};\citealt{Yoon&Langer04};\citealt{Piro08}), the neutron star formed during AIC is surrounded by a rotationally supported disk \citep{Dess06, Dess07,Ott10}.  For even modest differential rotation (that does not significantly change the Chandrasekhar mass), the disk mass is $M_D \sim 10^{-2}-0.1 \, M_\odot$.  However, if the WD is uniformly rotating, the disk mass is significantly lower, even if the WD is rotating near break-up \citep{Ott10}.  

Most of the rotationally supported material is initially just outside the proto-neutron star at a radius of $R_D \sim$ 100 km. The disk is born sufficiently hot that nuclei are dissociated into free nucleons and weak interaction processes have time to change the neutron to proton ratio. Under these conditions, the composition of the disk can be usefully described using the electron fraction
\begin{equation}
Y_e = \frac{n_p}{n_n+n_p}
\end{equation}
where $n_p$ and $n_n$ are the proton and neutron densities, respectively. Weak interactions in the disk are dominated by the pair capture reactions
\begin{equation}
e^+ + n \rightarrow \bar{\nu}_{e} + p
\label{eq:p}
\end{equation}
\begin{equation}
e^- + p \rightarrow \nu_{e} + n
\label{eq:n}
\end{equation}
where the resulting neutrino emission is also the dominant coolant for the disk material because the photons are trapped and cannot escape on the timescales of interest.  Although the disk is formed very hot, electron degeneracy suppresses the formation of electron/positron pairs.  The resulting electron excess favors electron captures (eq.~[\ref{eq:n}]) over positron captures (eq.~[\ref{eq:p}]), thus driving the composition neutron rich (e.g.~\citealt{Pruet+03}).

As the disk accretes onto the proto-neutron star, the remaining mass spreads to larger radii in order to conserve angular momentum.   As the disk spreads, the temperature drops (for the majority of the mass) and thus the neutrino cooling and weak reactions become less efficient.
When the bulk of the disk reaches $R_D \sim$ 1000 km, the disk has cooled to T $\sim 10^{10}$K, and weak interactions freeze out \citep{MPQ09-Neutron}.  Absent a central neutron star -- as is the case for analogous disks formed during binary neutron star  mergers -- weak interactions freeze out when the material is still quite neutron rich, with $Y_e \sim 0.2-0.4$.   During AIC, however, the proto-neutron star emits copious neutrinos in the first few seconds after formation \citep{BL}, which irradiate the disk as it spreads.  Because the neutron star is deleptonizing, electron neutrinos dominate the emission and the rate of neutrino captures $\nu_{e} + n \rightarrow p + e^{-}$ becomes comparable to the rate of pair captures (eqs.~[\ref{eq:p}]-[\ref{eq:n}]).  For a significant fraction of the mass, this increases the electron fraction at freeze-out to $Y_e \gtrsim$ 0.5.

As the disk continues to spread, the free nucleons in the disk recombine to form heavy nuclei 
once T $\sim$ few $\times 10^9$ K.   The formation of He alone releases $\sim 7$ MeV per nucleon, which is larger than the gravitational binding energy of the disk when fusion commences ($\sim 2$ MeV per nucleon).  As a result, the disk is unbound, with its slow viscous spreading transformed into a supersonic outflow that reaches $v \simeq 0.1$ c.  The end state of the fusion that initiates the disk's destruction is determined by the density, temperature, and electron fraction (the latter because weak interactions have frozen out); the parameters are such that the ejecta is nearly in nuclear statistical equilibrium (NSE).
Since there are roughly equal numbers of protons and neutrons, a significant fraction of the unbound disk fuses to form $^{56}$Ni,  the decay of which can power a supernova-like transient once the ejecta expands sufficiently that photons may diffuse out.  In the remainder of this paper, we calculate the lightcurves and spectra of this transient.

\vspace{-0.5cm}
\section{Assumptions \& Model Parameters}
\label{sec:fidmod} 

\subsection{Radiative Transfer Calculations}
\label{sec:fidmodcode}

We calculate lightcurves and spectra using the time-dependent radiative transfer code SEDONA, a detailed description of which can be found in \cite{KTN}.   SEDONA solves the radiative transfer problem in a homologously expanding atmosphere described by a velocity-time grid. We include the gamma ray energy emitted from the decay chain $^{56}$Ni $\rightarrow$ $^{56}$Co $\rightarrow$ $^{56}$Fe and evolve the composition accordingly.  We modified SEDONA to treat $^{56}$Ni as distinct, allowing us to include other Ni isotopes in the opacity without evolving their composition or including their decay energy. All line information is from \cite{Kurucz}.

SEDONA tracks the emitted gamma rays from radioactive decay and determines the resulting heating through Compton scattering and photoelectric absorption. The opacities are first calculated using the input density and composition, the calculated energy deposition, and an initial temperature guess.  The dominant opacities are electron scattering and bound-bound line transitions. Our calculations assume local thermodynamic equilibrium (LTE), although SEDONA supports non-LTE opacity calculations as well. The source function for the lines is calculated in the two-level atom formalism. To account for line scattering, the source function, $S_{\lambda}$, deviates from its LTE form of the Planck function, $B_{\lambda}$, through
\begin{equation}
S_{\lambda} = (1-\epsilon)J_{\lambda}+ \epsilon B_{\lambda}
\end{equation}
where the degree of line scattering is encapsulated in the thermalization parameter, $\epsilon$, which equals the ratio of absorption opacity to total opacity.  We fixed $\epsilon = 0.3$, motivated by the calculations of \citet{KTN}.  Given the opacity and ionization states in LTE, the temperature profile is determined from the condition of radiative equilibrium, balancing the rate of thermal emission with the rate of photon absorption and radioactive energy deposition.  The synthetic spectrum is determined using a Monte Carlo calculation.   The spectrum and temperature profile are iterated on until the temperature structure converges.   The final  synthetic lightcurves and spectra are determined on the last iteration by recording the photons that escape the outflow.   We set the velocity and time bin resolutions in our calculations to 200 km s$^{-1}$ and 0.1 day, respectively. 

\subsection{Outflow Properties}
\label{sec:fidmodics}

Although the outflows produced during AIC are likely to be asymmetric (albeit axisymmetric), we use a 1D spherically symmetric model to study the lightcurves and spectra.  Given current uncertainties in the properties of the unbound accretion disk and the interaction between the disk and the AIC supernova, we believe that this approximation is a reasonable first step.

The bulk speed of the outflow is $\simeq 0.1$ c, determined by the nuclear energy released as the disk is unbound.  Given the likely order unity fluctuations about this fiducial speed, we set the density distribution of the homologously expanding outflow to be
\begin{equation}
\rho=\begin{cases}
ae^{v/v_e} & \text{if $0 < v < 0.05 \, c$} \\
bv^{-C}     & \text{if $0.05 \, c  \leq v < 0.2 \, c $}
\end{cases}
\label{rhoi}
\end{equation}
where $C$ parameterizes the density profile of the outflow.  The constants $a$ and $b$, which in general are functions of time, are determined by continuity of density at $v = 0.05 \, c$ and by fixing the total outflow mass, $M_{tot}$.  The e-folding velocity, $v_e$, is included to ensure that there is not an artificially abrupt cutoff in the density at small velocity/radii (which can affect the photosphere at late times), but the exact value is not crucial; we take $v_e = 500 \, \kms$.  Equation (\ref{rhoi}) implies that the bulk of the mass in the outflow has $0.05c \le v < 0.2c$.  The precise value of the density power law index $C$ within this velocity range does not significantly affect the observational signatures of AIC.   We will focus on $C = 4$ but we carried out calculations for $C = 2-4$, which covers the full range of possibilities, from most of the mass being at high velocity ($C = 2$) to most of the mass being at smaller velocity ($C = 4$).  The lightcurves in the models with smaller $C$ peak somewhat earlier ($\Delta t = 0.2$ d between $C = 2$ and $C = 4$) and fall more rapidly after the peak, as would be expected since more of the mass is at high velocity.  Nonetheless, the peak bolometric luminosity remains nearly the same.  The spectra at early times are relatively independent of $C$, but there are some differences at t $\gtrsim$ 3 days when individual spectral features begin to become more prominent.   We will highlight these differences below, but we focus on the case of $C = 4$.

\begin{table}
\centering
\caption{Mass Fractions in the Three Fiducial Models.}
\begin{tabular}{c c c c}
\hline \hline
& & $Y_e$ range & \\
Element & 0.425-0.55 & 0.45-0.55 & 0.5-0.55 \\ \hline
Ca & $3.60\times10^{-5}$ & $4.55\times10^{-5}$ & $1.68\times10^{-5}$ \\
Sc & $1.58\times10^{-6}$ & $1.99\times10^{-6}$ & $3.62\times10^{-6}$ \\
Ti & $5.63\times10^{-4}$ & $2.55\times10^{-5}$ & $4.44\times10^{-5}$\\
V & $1.40\times10^{-6}$ & $1.77\times10^{-6}$ & $3.22\times10^{-6}$ \\
Cr & $2.24\times10^{-3}$ & $1.16\times10^{-4}$ & $7.61\times10^{-5}$ \\
Fe & $1.87\times10^{-2}$ & $1.12\times10^{-2}$ & $1.07\times10^{-2}$ \\
Co & $2.95\times10^{-3}$ & $3.73\times10^{-3}$ & $1.76\times10^{-3}$ \\
Ni & $8.59\times10^{-1}$ & $9.00\times10^{-1}$ & $9.87\times10^{-1}$ \\
Cu & $5.03\times10^{-3}$ & $4.52\times10^{-3}$ & $2.21\times10^{-4}$ \\
Zn & $1.03\times10^{-1}$ & $7.31\times10^{-2}$ & $5.90\times10^{-5}$ \\
Ga & $1.42\times10^{-4}$ & $1.40\times10^{-4}$ & $0.00$ \\
Ge & $3.86\times10^{-3}$ & $4.23\times10^{-3}$ & $0.00$ \\
As & $5.60\times10^{-6}$ & $7.08\times10^{-6}$ & $0.00$ \\
Se & $5.97\times10^{-4}$ & $5.42\times10^{-4}$ & $0.00$ \\
Rb & $1.69\times10^{-4}$ & $1.25\times10^{-5}$ & $0.00$ \\
Sr & $2.86\times10^{-3}$ & $1.12\times10^{-3}$ & $0.00$ \\
Y & $1.84\times10^{-4}$ & $2.19\times10^{-4}$ & $0.00$ \\
Zr & $9.46\times10^{-4}$ & $1.20\times10^{-3}$ & $0.00$ \\
Mo & $3.40\times10^{-6}$ & $4.30\times10^{-6}$ & $0.00$ \\
\end{tabular}
\label{table:comp_disk}
\end{table}

\begin{table}
\centering
\caption{Mass Fraction of $^{56}$Ni in the Three Fiducial Models.}
\begin{tabular}{c c c c}
\hline \hline
& & $Y_e$ range & \\
& 0.425-0.55 & 0.45-0.55 & 0.5-0.55 \\ \hline
$\frac{X_{^{56}Ni}}{X_{Ni}}$ & $5.38\times10^{-1}$ & $6.49\times10^{-1}$ & $9.97\times10^{-1}$ \\
\end{tabular}
\label{table:ni56_disk}
\end{table}

\cite{MPQ09-Nickel} showed that different models for the evolution of the accretion disk around the NS (e.g., different values of the Shakura-Sunyaev 1973 viscosity parameter $\alpha$) can lead to somewhat different composition in the resulting ejecta.   Specifically, the electron fraction is within the range  $0.4 < Y_e < 0.65$, but the distribution of mass as a function of $Y_e$ depends on the details of the disk's evolution. To cover this parameter space we studied three fiducial models which include an increasingly large amount of neutron-rich material at low $Y_e$:  (1) $Y_e = 0.5-0.55$, (2) $Y_e = 0.45-0.55$ and (3) $Y_e = 0.425-0.55$.   The models all have equal mass per unit $Y_e$.   We believe that the models with the larger range of electron fraction (in particular, $Y_e = 0.425-0.55$) are the most realistic because the disk models in \cite{MPQ09-Nickel} in general have a range of $Y_e$ at different radii and times.   However, the differences between these three different models give an indication of the systematic uncertainties in our predictions introduced by uncertainties in the outflow composition.  Lacking a hydrodynamic calculation of the disk evolution and its transition to the ejecta phase, we take the composition of the ejecta to be uniform at all radii/velocites.

\begin{figure*}
\centering
\subfigure[Bolometric Lightcurves]{
\begin{overpic}[scale=0.3]{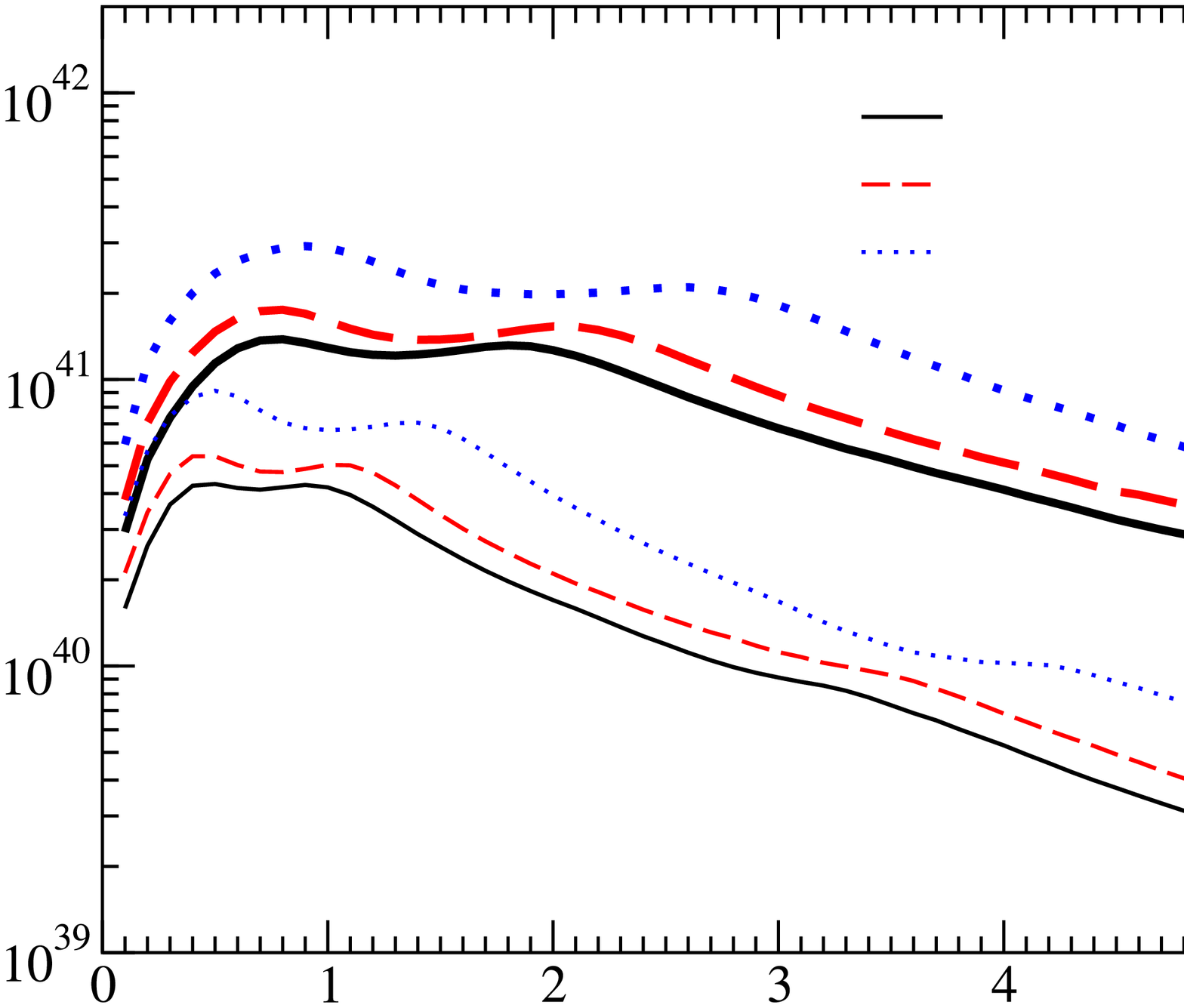}
\put(113,15){Time (d)}
\put(8,40){\begin{sideways}Bolometric Luminosity $\left(\textrm{ergs s}^{-1}\right)$\end{sideways}}
\put(153,149){$Y_e : 0.425 - 0.55$}
\put(153,139.5){$Y_e : 0.45 - 0.55$}
\put(153,130){$Y_e : 0.5 - 0.55$}
\put(60,140){$M_{tot} = 10^{-2} M_{\odot}$}
\put(60,60){$M_{tot} = 3\times 10^{-3}M_{\sun}$}
\end{overpic}
}
\subfigure[R-Band Lightcurves]{
\begin{overpic}[scale=0.3]{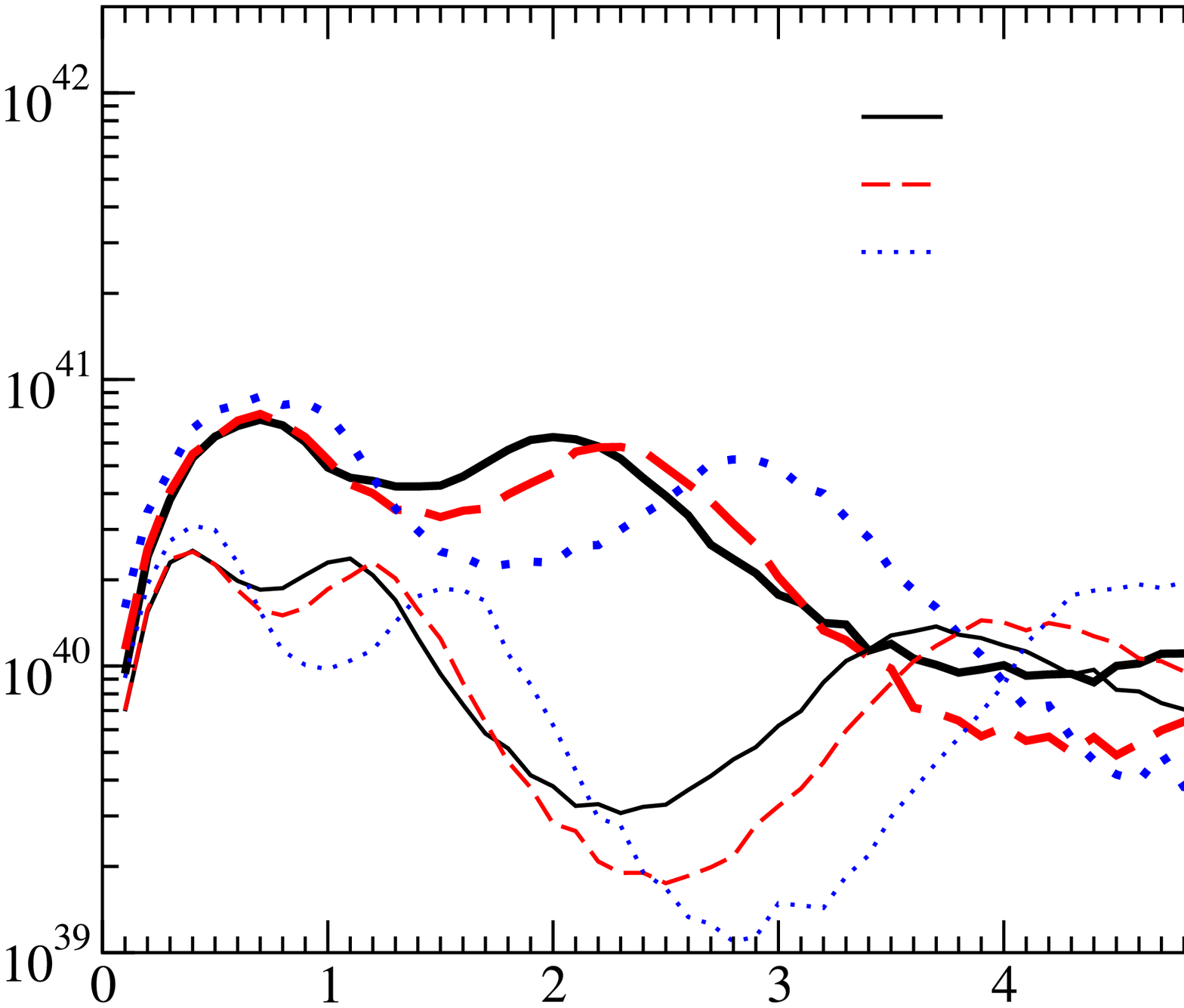}
\put(113,15){Time (d)}
\put(8,40){\begin{sideways}R-Band Luminosity $\left(\textrm{ergs s}^{-1}\right)$\end{sideways}}
\put(153,149){$Y_e : 0.425 - 0.55$}
\put(153,139.5){$Y_e : 0.45 - 0.55$}
\put(153,130){$Y_e : 0.5 - 0.55$}
\put(40,122){$M_{tot} = 10^{-2} M_{\odot}$}
\put(40,42){$M_{tot} = 3\times 10^{-3}M_{\sun}$}
\end{overpic}
}
\caption{Predicted bolometric and R-band lightcurves of AIC transients, calculated for different values of the total ejected mass $M_{tot}$.  The R-band center was taken as $\lambda_{\textrm{R}} = 7011 \textrm{ \AA}$.  Thick and thin lines correspond to $M_{tot} = 10^{-2} M_{\odot}$ and $3\times 10^{-3}M_{\sun}$, respectively.  Different linestyles correspond to different ejecta compositions, specifically different assumptions about the range of electron fraction in the ejecta (see Tables \ref{table:comp_disk} and \ref{table:ni56_disk}).}
\label{fig:bol-lightcurves}
\end{figure*}

Given the electron fraction in the ejecta, we determined the composition of each model using the nucleosynthesis calculations of \cite{WH} and \cite{Seit}, averaging over the range of $Y_e$ in our model, and using the solar system abundances from \cite{Arnett} to determine absolute abundances.   Table \ref{table:comp_disk} summarizes the composition of each of the models and Table \ref{table:ni56_disk} gives the mass fraction of $^{56}$Ni in each model.  In determining the composition of our models, we used the calculations of  \cite{WH} and \cite{Seit} with thermodynamic properties most similar to AIC ejecta.  Specifically, for $0.425 \leq Y_e \leq 0.5$, we used the finite entropy abundances calculated from the $\alpha$-process with $T = 10^{10} K$ and $\rho = 2 \times 10^7$g cm$^{-3}$ in Table 3 of \cite{WH}, corresponding to a mild $\alpha$-rich freeze out. The entropy and expansion time scale used in these calculations are similar to those expected in the disk when $\alpha-$particles form and outflows begin.  For $0.5 < Y_e \leq 0.55$, we use the abundances calculated from NSE with $T = 3.5 \times 10^9 K$ and $\rho = 10^7$ g cm$^{-3}$ in Figure 3 of \cite{Seit}.  Because AIC outflows likely possess a moderate entropy $\sim 3-10 \, k_B$ baryon$^{-1}$ \citep{MPQ09-Nickel}, and hence may undergo a mild $\alpha-$rich freeze-out, the pure NSE calculations in \cite{Seit} do not stricly apply.  However, by extrapolating the results of full nucleosynthesis calculations by \citet{Pruet+04} to the relevant entropy regime, we find that the yield of $^{56}$Ni, the dominant isotope in this composition range, is well-approximated by its NSE abundance.

Since the nuclei are in their post-decay phase in \cite{WH}, we converted the entire fraction of $^{56}$Fe to its parent $^{56}$Ni.   Although some matter with $Y_e < 0.425$ may be present in AIC ejecta, we did not include it in our models due to the uncertain and contradictory nucleosynthesis information in \cite{Seit} and \cite{HWE}. We also did not include Br and Kr in our calculations due to lack of line information in \cite{Kurucz}.  Finally, our calculations only include heating due to the decay of $^{56}$Ni, although other radioactive isotopes are produced and may contribute to the light curve.  In particular, we find that $^{57}$Ni (half-life $\tau_{1/2} = 35.6$ hr) and $^{62}$Zn ($\tau_{1/2} = 9.2$ hr) produce {\it specific} heating rates that are a factor $\sim 6$ and $\sim 12$ times higher, respectively, than $^{56}$Ni on timescales $t \lesssim \tau_{1/2}$.  However, in most cases the much lower mass fractions of these isotopes relative to $^{56}$Ni implies that their contribution to the total heating is relatively minor.  One exception is the mass fraction of $^{62}$Zn in our $Y_{e} = 0.425-0.55$ model, which we estimate is only a factor $\sim 4$ less than $^{56}$Ni and thus could contribute appreciably to the very early light curve.  

\begin{figure*}
\centering
\subfigure[U-B]{
\begin{overpic}[scale=0.3]{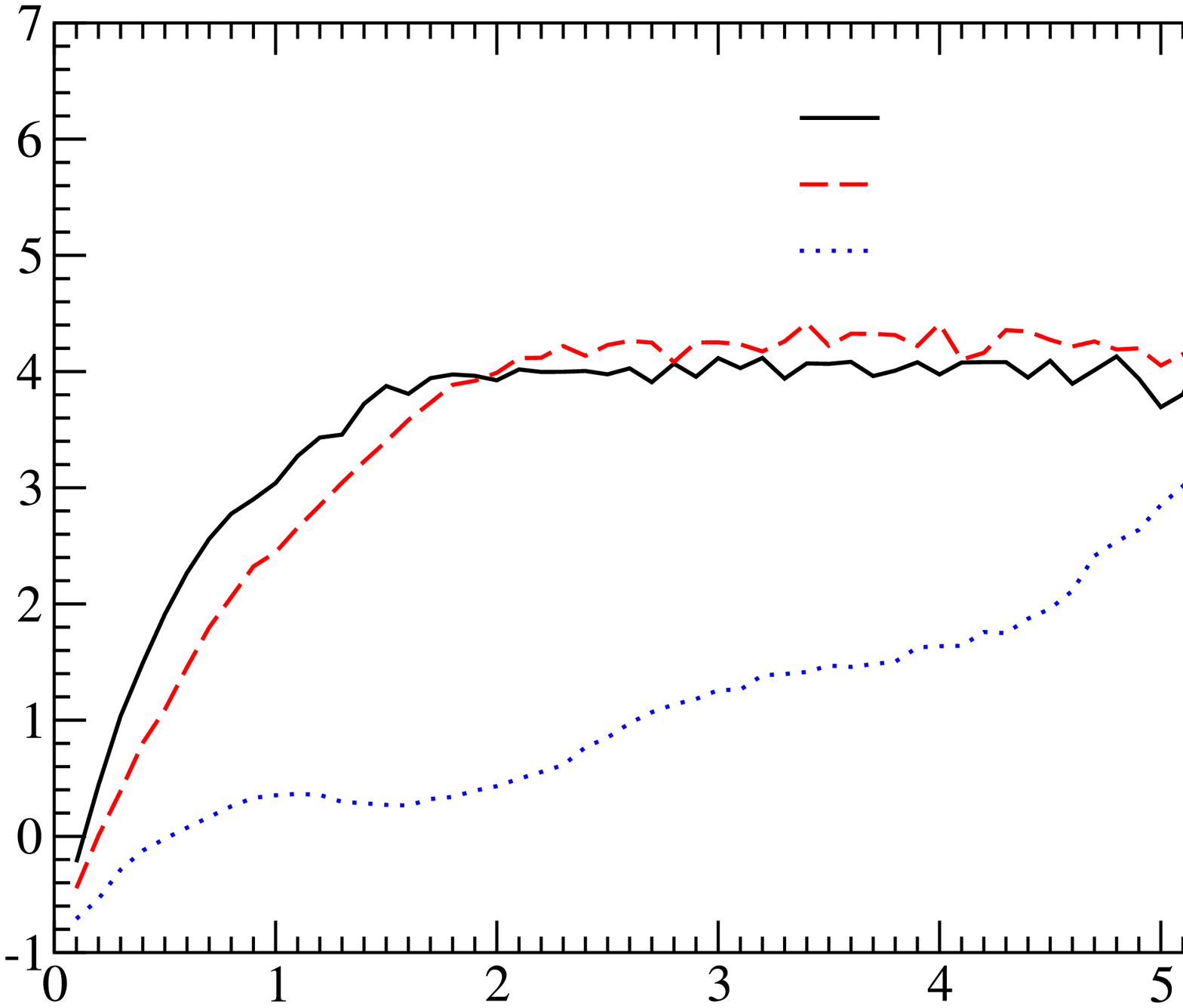}
\put(104,5){Time (d)}
\put(8,80){\begin{sideways}U-B\end{sideways}}
\put(144,141.5){$Y_e : 0.425 - 0.55$}
\put(144,132){$Y_e : 0.45 - 0.55$}
\put(144,122.5){$Y_e : 0.5 - 0.55$}
\end{overpic}
}
\subfigure[B-V]{
\begin{overpic}[scale=0.30]{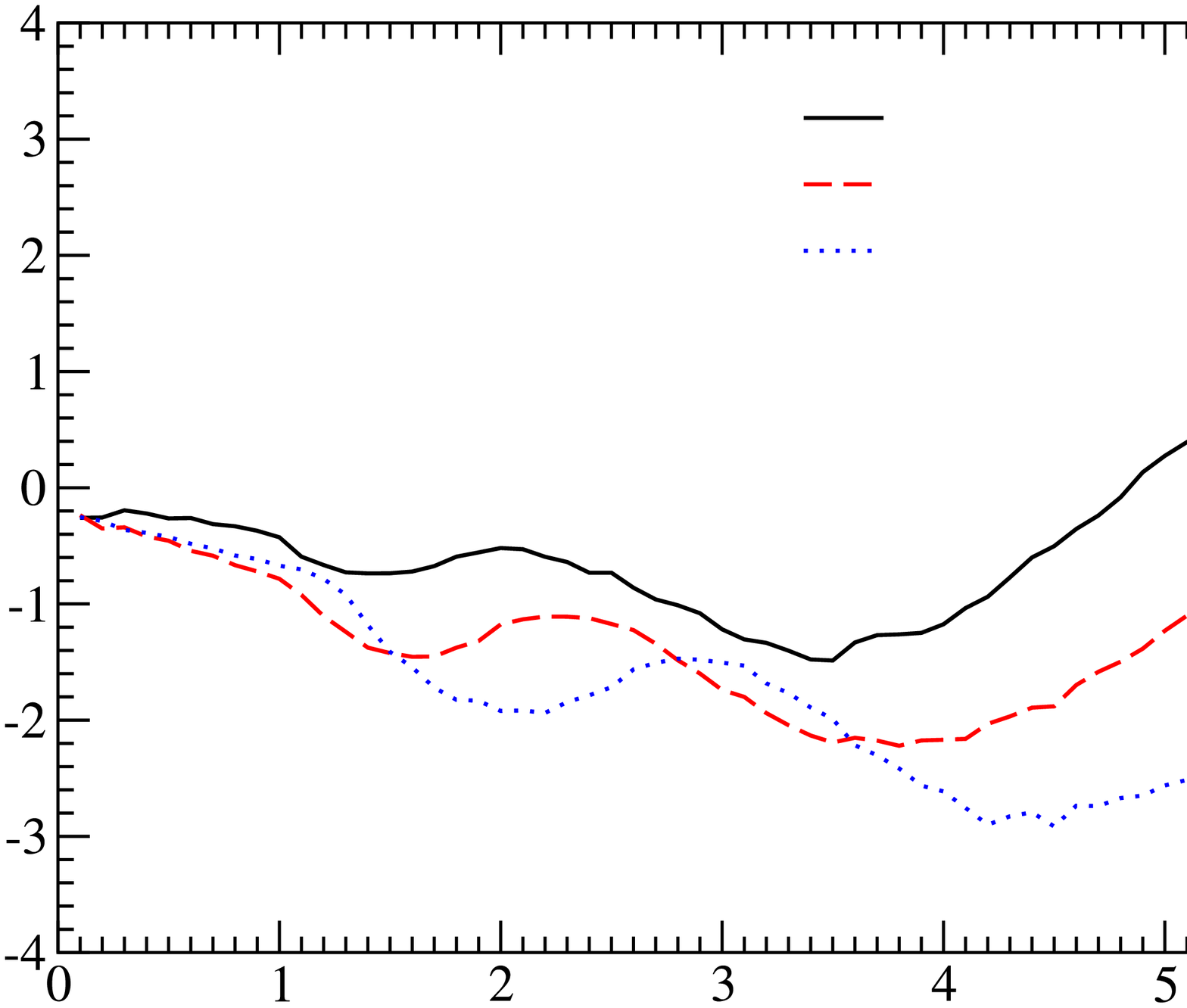}
\put(104,5){Time (d)}
\put(8,80){\begin{sideways}B-V\end{sideways}}
\put(144,141.5){$Y_e : 0.425 - 0.55$}
\put(144,132){$Y_e : 0.45 - 0.55$}
\put(144,122.5){$Y_e : 0.5 - 0.55$}
\end{overpic}
}
\subfigure[B-J]{
\begin{overpic}[scale=0.30]{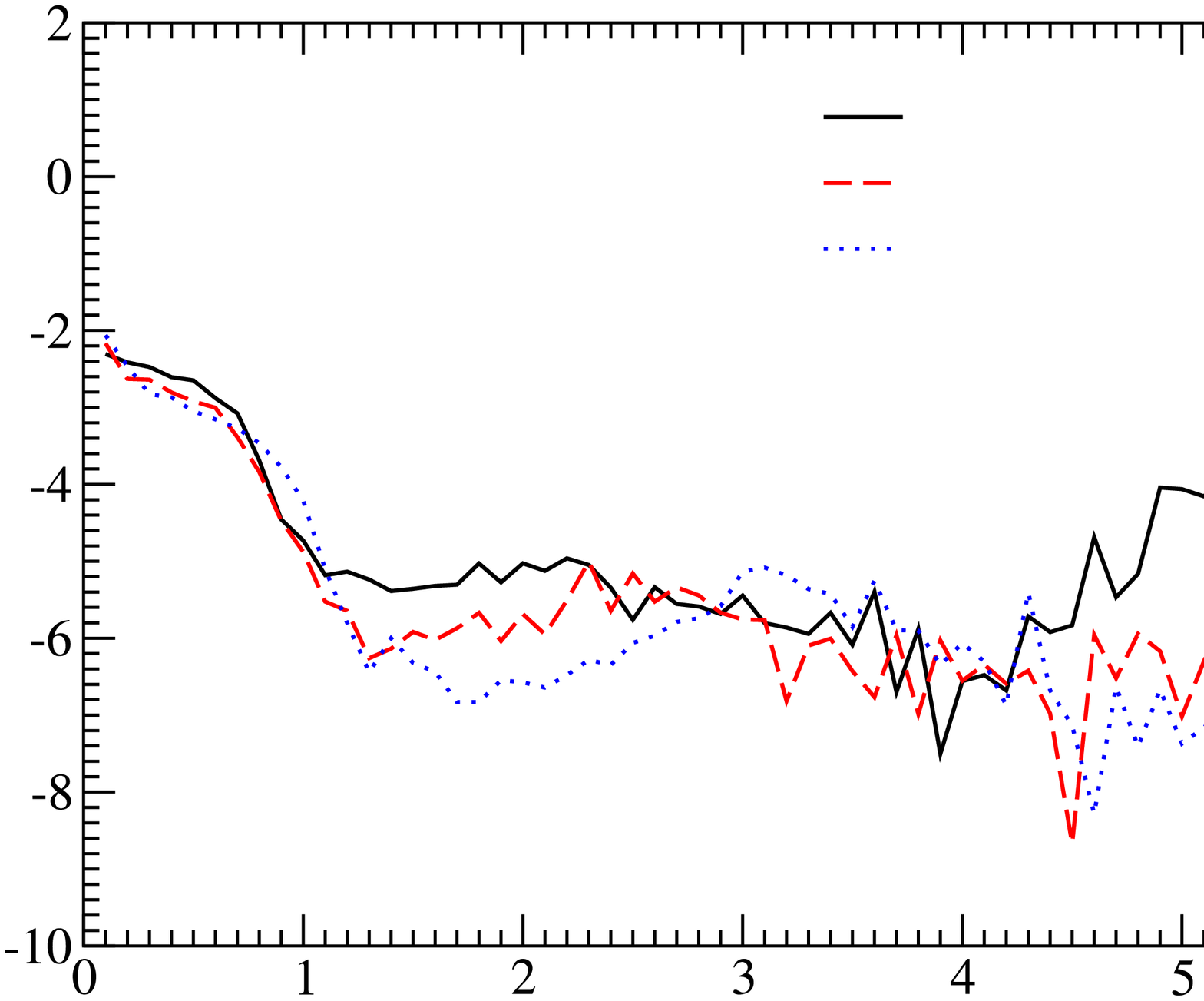}
\put(104,5){Time (d)}
\put(8,80){\begin{sideways}B-J\end{sideways}}
\put(144,141.5){$Y_e : 0.425 - 0.55$}
\put(144,132){$Y_e : 0.45 - 0.55$}
\put(144,122.5){$Y_e : 0.5 - 0.55$}
\end{overpic}
}
\subfigure[R-I]{
\begin{overpic}[scale=0.30]{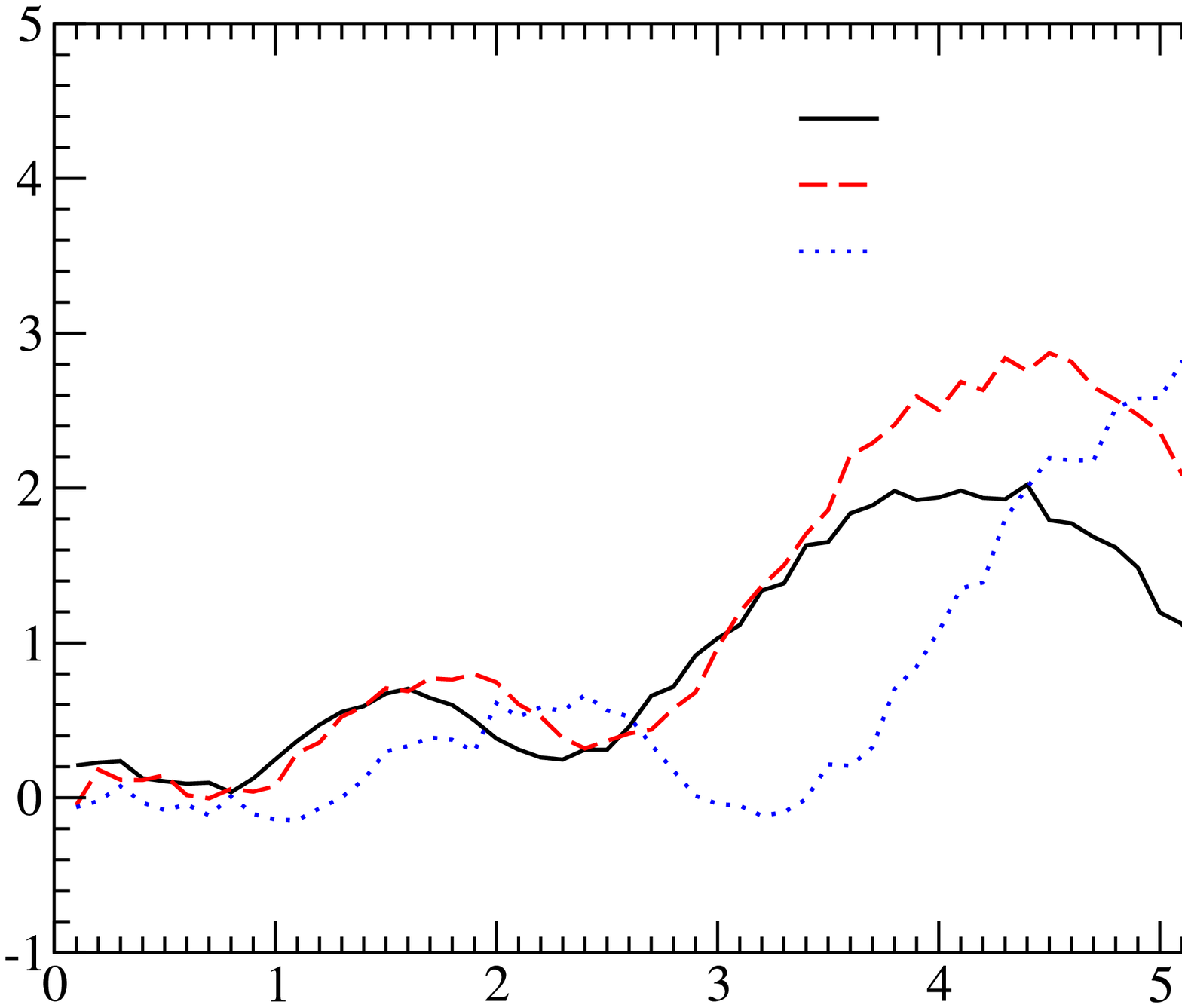}
\put(104,5){Time (d)}
\put(8,80){\begin{sideways}R-I\end{sideways}}
\put(144,141.5){$Y_e : 0.425 - 0.55$}
\put(144,132){$Y_e : 0.45 - 0.55$}
\put(144,122.5){$Y_e : 0.5 - 0.55$}
\end{overpic}
}
\caption{U-B, B-V, B-J, and R-I color evolution for the model with ejecta mass $M_{tot} = 10^{-2}M_{\sun}$.   The models with the wider range of electron fraction are, we believe, more physical ($Y_e = 0.425-0.55$ and $Y_e = 0.45-0.55$); see text.   These initially become redder with time in U-B, with little evolution after a few days.   The spectra at 1 and 5 days are shown in Figure \ref{fig:spectra}.  The colors are defined using the ratio of the luminosities $\lambda L_\lambda$ at the following wavelengths:  $\lambda_{\textrm{U}} = 3611 \textrm{ \AA}, \lambda_{\textrm{B}} = 4611 \textrm{ \AA}, \lambda_{\textrm{V}} = 5851 \textrm{ \AA}, \lambda_{\textrm{R}} = 7011 \textrm{ \AA}, \lambda_{\textrm{I}} = 8111 \textrm{ \AA}$, and $\lambda_{\textrm{J}} = 12351 \textrm{ \AA}$ (chosen to match the wavelength bins used in the radiative transfer calculation).}
\label{fig:color-lightcurves}
\end{figure*}

As discussed in \S \ref{sec:summary}, the total mass in the ejecta produced during AIC is uncertain, and depends on the differential rotation in the WD progenitor.    \cite{MPQ09-Nickel} determined the mass of unbound disk material to be a few times $10^{-2}M_{\odot}$ using collapse calculations from \cite{Dess06, Dess07}, but the ejecta mass could be significantly lower if the WD is more uniformly rotating \citep{Ott10}.   We studied a range of total masses for the ejecta but focus on $M_{tot} = 3 \times 10^{-3} M_\odot$ and $M_{tot}=10^{-2}M_{\odot}$ here because this is the range of masses for which the disk ejecta produces an observationally interesting luminosity.  The lightcurves and spectra have similar behaviour for different total masses, simply shifted to somewhat earlier times and shorter wavelengths for lower total mass, as would be expected using analytic estimates based on Arnett's law (\citealt{Arnett82}).

\vspace{-0.8cm}
\section{Results}
\label{sec:results}

Figure \ref{fig:bol-lightcurves} shows the bolometric and R-band lightcurves produced by AIC ejecta for the three different compositions summarized in Tables  \ref{table:comp_disk} and  \ref{table:ni56_disk}, and for two different total masses.   The lightcurve evolution is consistent with analytic expectations based on the competition between expansion and diffusion (\citealt{Arnett82}).   The low mass and high speed of the ejecta imply that the rise time is extremely fast $\sim$ day; the subsequent decay is somewhat slower, with the luminosity decaying significantly after $\simeq 4-5$ days for $M_{tot} = 10^{-2} M_\odot$ and $\simeq 2-3$ days for $M_{tot} =  3 \times10^{-3} M_\odot$.    The peak bolometric luminosity is $\simeq 1-3 \times 10^{41} \, \ergss$ for $M_{tot} = 10^{-2} \, M_\odot$ and a factor of $\simeq 3$ lower for $M_{tot} = 3 \times 10^{-3} \, M_\odot$.   This is a factor of $\sim 100$ fainter than a typical Type Ia supernova, which is consistent with the much smaller Ni masses produced during AIC.

For a given total mass, Figure \ref{fig:bol-lightcurves} shows that the models with smallest range in electron fraction ($Y_e = 0.5-0.55$) are the brightest and broadest.  The dependence on the composition is, however, modest and is reasonably degenerate with a different total mass for the ejecta.    
This implies that the uncertainty in the detectability of AIC is due primarily to uncertainty in the total mass of the disk ejecta, not the composition of the material.  The dependence on $Y_e$ in Figure \ref{fig:bol-lightcurves} arises because only matter with $Y_e \gtrsim 0.48$ produces significant Ni, which both powers the lightcurve and contributes to the opacity of the ejecta.   Lower $Y_e$ material ends up as higher Z elements which do not contribute to the radioactive heating (note that for even lower $Y_e \simeq 0.1-0.3$, the decay of neutron-rich ejecta can produce a heating rate similar to that of Ni; \citealt{Metzger10}).

The bumps in the AIC lightcurves in Figure \ref{fig:bol-lightcurves} have the same physical cause as those seen in some Type Ia SNe (see \citealt{pinto2000} and \citealt{kasen06} for a discussion of the latter).   As the ejecta expands, the matter cools, and ions begin to recombine.  Singly ionized ions have a greater number of lines in the optical and infrared part of the spectrum, and are therefore more efficient at fluorescing photons from bluer to redder wavelengths.  Because the ejecta is more transparent at redder wavelengths, this redistribution effect allows the radiation energy stored in the ejecta to escape more rapidly, leading to a rise in the bolometric light curve.

Figure \ref{fig:color-lightcurves} shows the U-B, B-V, B-J, and R-I colors as a function of time for the $M_{tot} = 10^{-2} M_\odot$ models from Figure \ref{fig:bol-lightcurves}.  We define the colors here using the ratio of the luminosities $\lambda L_\lambda$ at the centers of the appropriate bands, with $\lambda_{\textrm{U}} = 3611 \textrm{ \AA}, \lambda_{\textrm{B}} = 4611 \textrm{ \AA}, \lambda_{\textrm{V}} = 5851 \textrm{ \AA}, \lambda_{\textrm{R}} = 7011 \textrm{ \AA}, \lambda_{\textrm{I}} = 8111 \textrm{ \AA}$, and $\lambda_{\textrm{J}} = 12351 \textrm{ \AA}$.   Figure \ref{fig:color-lightcurves} shows that AIC is predicted to become increasingly red in the first $\sim$ day, after which there is not as much color evolution as a function of time.  This is particularly true for our preferred models which have a wider range of electron fraction in the outflow.   The model with the narrow range of $Y_e$ (0.5-0.55) is bluer and shows somewhat more color evolution at early times, as a result of being almost entirely $^{56}$Ni (Tables  \ref{table:comp_disk} and  \ref{table:ni56_disk}).   Note, however, that although the emission as a whole becomes redder with time, the optical colors are slightly blue, with $B - V \sim -1$ to $-2$.

The lower mass models with $M_{tot} = 3 \times 10^{-3}M_{\odot}$ show the same qualitative U-B color evolution as the $M_{tot} = 10^{-2}M_{\odot}$ case.  In particular, they reach the same color plateau, but somewhat faster than in the high mass case because the ejecta becomes transparent earlier.  One difference, however, is that for $M_{tot} = 3\times 10^{-3}M_{\odot}$, the B-V and B-J colors become redder at day 3, and the R-I colors peak at the same value 2 days earlier than the high mass cases and rise and fall faster.  Models with shallower density profiles (C = 3 and 2 in eq.~[\ref{rhoi}]) show the same qualitative behavior for U-B, B-V, and U-J as our standard C = 4 model;  they are, however, slightly bluer in U-B throughout (by $\sim 0.6, 1.2$), respectively, while in B-V and B-J they are equal until day 3, after which they overturn and become redder by $\sim 1, 2$ magnitude, respectively. In R-I, they peak at the same value but $\sim 0.9, 1.5$ days earlier, respectively, and rise and fall faster.

Our predicted color evolution may be sensitive to non-LTE effects.  In particular, the red colors are primarily due to the fact that in these models a significant fraction of the mass is in iron and nickel even at the highest velocities.  The outer layers of the ejecta are cooler, and in a lower ionization state, and so they very strongly  blanket the bluer wavelengths.   Non-thermal ionization from the decay products of $^{56}$Ni could in principle keep the outer layers of the ejecta more ionized.  This would reduce the degree of line blanketing in the blue part of the spectrum (and the fluorescence to redder wavelengths) leading to an overall less red spectrum.   This possibility will be studied in future work.

\begin{figure*}
\centering
\subfigure[t = 1 day]{
\begin{overpic}[scale=0.30]{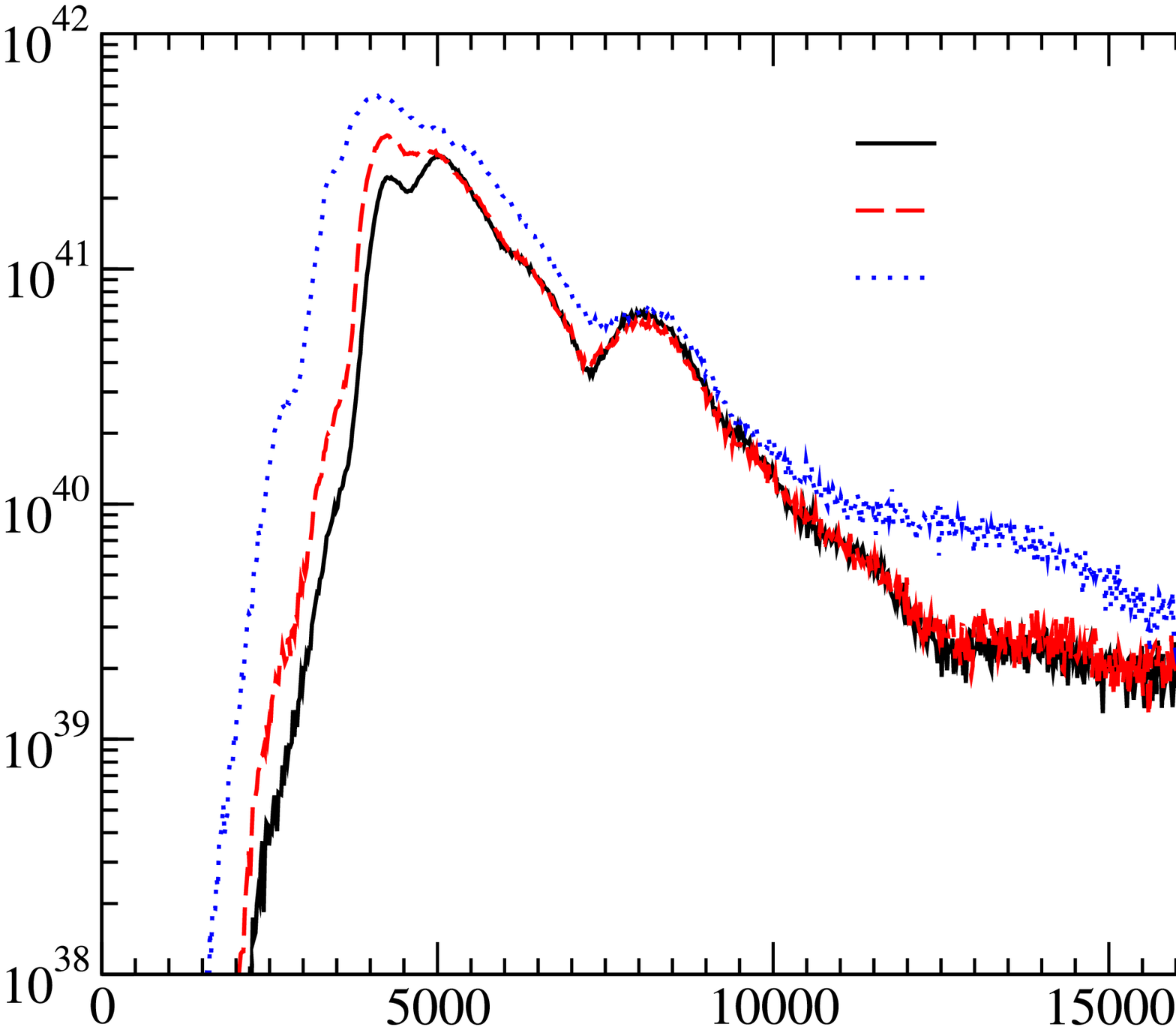}
\put(90,15){Wavelength $\left(\textrm{\AA}\right)$}
\put(8,70){\begin{sideways}$\lambda L_{\lambda}$ $\left(\textrm{ergs s}^{-1}\right)$\end{sideways}}
\put(153,149){$Y_e : 0.425 - 0.55$}
\put(153,139.5){$Y_e : 0.45 - 0.55$}
\put(153,130){$Y_e : 0.5 - 0.55$}
\end{overpic}
}
\subfigure[t = 5 day]{
\begin{overpic}[scale=0.30]{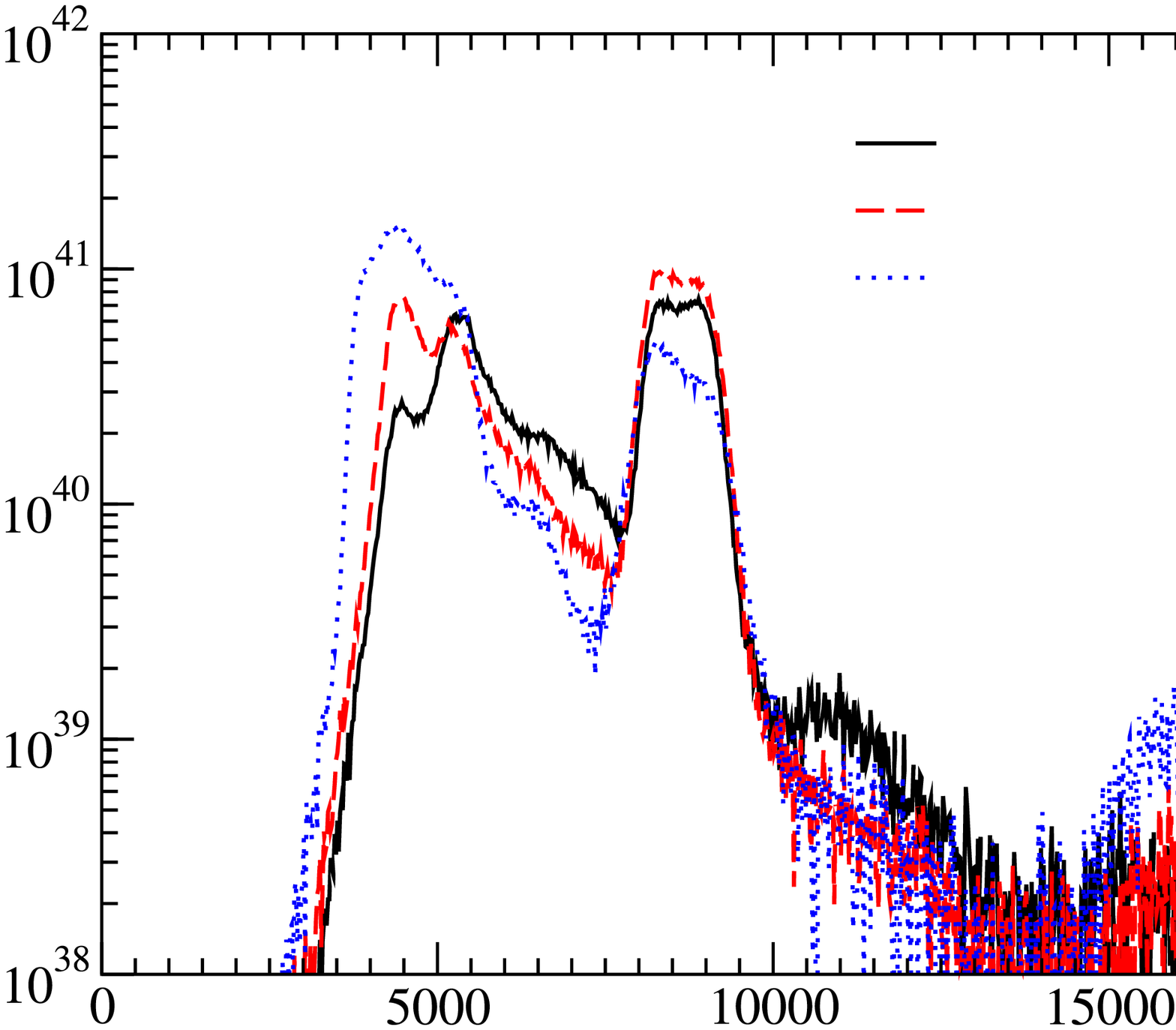}
\put(90,15){Wavelength $\left(\textrm{\AA}\right)$}
\put(8,70){\begin{sideways}$\lambda L_{\lambda}$ $\left(\textrm{ergs s}^{-1}\right)$\end{sideways}}
\put(153,149){$Y_e : 0.425 - 0.55$}
\put(153,139.5){$Y_e : 0.45 - 0.55$}
\put(153,130){$Y_e : 0.5 - 0.55$}
\end{overpic}
}
\caption{Spectra for the model with ejecta mass $M_{tot} = 10^{-2}M_{\sun}$.  At early times the spectra are dominated by Doppler-broadened Nickel features, with no individual spectral features identifiable.  At 5 days, however, the Ca triplet at 9000 \AA \ is quite prominent.   This line is generically present at late times ($\simeq 2-5$ days) in all of our models.}
\label{fig:spectra}
\end{figure*}

The spectra at 1 day and 5 days are shown in Figure \ref{fig:spectra} for the $M_{tot} = 10^{-2} M_\odot$ models.   Near the peak of the lightcurve ($\sim 1-3$ days), the emission peaks at $\simeq 3000-6000$ \AA \enspace and is reasonably red, consistent with Figure \ref{fig:color-lightcurves}.   At 1 day, individual spectral features are not discernible:   given the high speeds, the lines have Doppler broadened and merged.  Spectral lines do become more apparent at later times.   In particular, at $\simeq 3-5$ days the calcium triplet at $\sim 9000$ \AA \enspace produces a luminosity comparable to the total luminosity of the ejecta, despite the fact that the mass fraction of Ca is $\sim 3 \times 10^{-5}$.  This is the most prominent individual spectral feature seen in any of our calculations, although we note that the actual strength of the Ca II line may be somewhat overestimated due to our LTE assumption.   The spectra are generally Ni dominated but Doppler broadening makes it very difficult to identify individual spectral features.   Although the results in Figure \ref{fig:spectra} are for $M_{tot} = 10^{-2} M_\odot$ and $\rho \propto v^{-4}$ in equation (\ref{rhoi}), the general conclusions about the spectra of AIC drawn here are reasonably robust and also apply to the other calculations we have carried out.    In particular, the lower mass models with $M_{tot} = 3 \times 10^{-3}M_{\sun}$ have similar spectra at day 1.    Models with different density profiles have similar spectra at day 1, but steeper velocity profiles (larger C) retain the Ca II triplet line and the peak at $\sim 5000$ \AA\ to later times, simply because more mass is concentrated at lower velocities. 

To assess the potential impact of the weak SN accompanying AIC on our predicted lightcurves and spectra, we also carried out a calculation in which we included the likely nucleosynthetic products of such a SN in our radiative transfer calculation.  We assume that the SN material is uniformly mixed with the disk wind ejecta, which ignores the potentially interesting dynamics of the interaction between the disk wind and the SN.  We set the mass in the SN ejecta to be $3\times10^{-3}M_{\odot}$, motivated by hydrodynamic simulations in \cite{Dess06} -- note that since there is no significant outer stellar envelope in AIC, the ``SN" mass is largely the mass unbound by the neutrino-driven wind from the proto-neutron star in the first $\sim 1$ sec after collapse.  Given that most of this ejecta has $Y_e \simeq 0.25-0.4$ \citep{Dess06}, it will primarily form elements near the second $r-$process peak at $A \sim 130$ \citep{Hoffman+97}.  Although a detailed nucleosyntheis calculation would be necessary to determine the wind's precise nuclear yield, for simplicity we include in our calculation elements in the range of the second peak (Z = 45-60) with an abundance distribution normalized to the observed solar values (taken from \citealt{Arnett}).  We omitted I and Xe because no line information is available in \cite{Kurucz}.  The resulting model has a slightly larger peak luminosity and rise time as compared to our fiducial model without the additional matter (shown in Fig. \ref{fig:bol-lightcurves}), but the lightcurve is somewhat broader and has a slower decay.  This is simply a consequence of the larger total mass of the ejecta.  Although the AIC plus SN wind case may be somewhat redder than AIC alone at late times, we did not find any signature (e.g., spectroscopic) that could distinguish entrained SN material from a disk wind with a somewhat larger mass.  We note, however, that the available line data on very heavy elements is incomplete, which may partially explain the lack of discernable spectroscopic features. 

\vspace{-0.6cm}
\section{Discussion}
\label{sec:discussion}

Although the theoretical possibility of AIC has been recognized for decades (\citealt{Canal&Schatzman76};\citealt{Nomoto+79}), only recently have transient surveys had sufficient depth, field of view, and cadence to potentially detect and characterize these events.  The identification of AIC, or the ability to place meaningful constraints on its rate, would have important implications for a variety of astrophysical problems, including degenerate binary evolution, the progenitors of Type Ia SNe, and the formation channels of globular cluster NSs \citep{Grindlay87}, low mass X-ray binaries \citep{vandenHeuvel84}, and millisecond pulsars \citep{Grindlay&Bailyn88}.  

Because AIC transients are powered by the decay of $^{56}$Ni, they have many qualitative similarities to Type Ia SNe.  Although the observed lightcurves are too dim to perform cosmology, characterizing the correlation between peak luminosity and decay timescale (i.e., the \citealt{Phillips93} relation) for AIC would be useful for observationally identifying these events.  For example, the duration and peak luminosity of AIC transients increases with the ejecta mass, which probably increases in proportion to the initial mass of the accretion disk \citep{MPQ09-Nickel}.  The disk mass, in turn, is set by the angular momentum and rotational profile of the WD prior to collapse (\citealt{Dess06}; \citealt{Ott10}), which also determines whether AIC is expected to be a powerful source of gravitational waves (\citealt{Ott09}).  Everything else being equal, more luminous AIC transients are thus more likely to be accompanied by a detectable gravitational waves, making AIC transients a promising electromagnetic counterpart to events detected with ground-based interferometers such as Advanced LIGO.  

We now briefly address the prospects for the ``blind'' detection of AIC with current and future transient surveys.  Although the rate of AIC is uncertain, the amount of rare elements produced by the neutron-rich SN ejecta constrains the rate $R_{\rm AIC}$ to be $\lesssim 10^{-4}$ yr$^{-1}$ per Milky Way-type galaxy (or $\lesssim 10^{-2}$ of the SN Ia rate) if the AIC calculations of \citet{Dess06} are representative.  Assuming that the AIC rate is proportional to the blue stellar luminosity \citep{Kopparapu+08}, a Galactic rate of $R_{\rm AIC} \equiv 10^{-4}R_{-4}$ yr$^{-1}$ corresponds to a volumetric rate of $10^{-6}R_{-4}$ Mpc$^{-3}$ yr$^{-1}$.  For an ejecta mass of $M_{\rm ej} = 10^{-2}M_{\sun}$, we find that AIC results in an optical transient with a peak luminosity $L_{\rm peak} =$ few $\times 10^{41}$ ergs s$^{-1}$ (Fig.~\ref{fig:bol-lightcurves}).  The resulting luminosity distance is 600(400)[100]$L_{41}^{1/2}$ Mpc for a magnitude limit of 25(24)[21], where $L_{41} \equiv L_{\rm peak}/10^{41}$ ergs s$^{-1}$.  The PanSTARRs Medium Deep Survey (MDS; \citealt{Kaiser+02}), which covers $\sim 84$ deg$^{2}$ in $g$ and $r$ down to AB magnitude $\sim 25$, should thus detect $\sim 2 L_{41}^{3/2}R_{-4}$ AIC transients per yr.  The 5-day cadence survey of the Palomar Transient Factory (PTF; \citealt{Law+09}), which surveys an active area $\sim 2700$ deg$^{2}$ to a limiting AB magnitude of 21, will detect $0.3 L_{41}^{3/2}R_{-4}$ yr$^{-1}$.  Thus, for $R_{\rm AIC} \sim 10^{-4}$ yr$^{-1}$ and $L_{41} \sim$ few we conclude that MDS and PTF should detect $\sim 1-3$ AIC per year.  As we discuss below, identifying AIC transients for follow-up observations will be challenging, so these estimates are probably optimistic and motivate faster cadence surveys such as the Palomar P60-FasTING survey \citep{Kasliwal+10}.  Prospects for detection are better with the Large Synoptic Survey Telescope (LSST; \citealt{Strauss+:2010}), which will image the entire sky down to a limiting magnitude $\sim 24.5$ every 3$-$4 nights and should detect AIC events at a rate $\sim 10^3 \, L_{41}^{3/2}R_{-4}$ yr$^{-1}$.    Unfortunately, the fraction of AIC that have ejecta masses of $\sim 10^{-2} M_\odot$, and thus peak luminosities of $L_{\rm peak} =$ few $\times 10^{41}$ ergs s$^{-1}$, is very difficult to predict given uncertainties in the differential rotation of the progenitor WD.   Absent additional theoretical progress, it will thus not be easy to turn upper limits from transient surveys into constraints on the rate of AIC.

Although the prospects that upcoming transient surveys will detect AIC are promising, their identification will be challenging due to the deluge of other transients that will be discovered (and, indeed, are already being discovered).  Other thermal transients predicted to occur on $\sim$ few day timescales include the failed deflagrations of C/O WDs \citep{Livne+05}, ``.Ia'' SNe due to unstable thermonuclear He flashes from WD binaries (\citealt{Bildsten+07}; \citealt{Shen+10}), shock breakout from Type 1a supernova \citep{piro2010}, and radioactively-powered transients from NS mergers (\citealt{Metzger10}).  

Although NS mergers and other WD-related explosions may originate from a similar stellar population as AIC, a defining characteristic of AIC is that the majority of the ejected material is processed through NSE.  This means that few low or intermediate-mass elements are likely to be produced (such as He, O, C, and Ti), in contrast to what is expected from .Ia's or the failed detonation of a WD.  We note, however, that trace amounts of Ca and Ti are produced in some of our models (see Table \ref{table:comp_disk}), which still may produce strong lines despite their low abundances (e.g.~the Ca triplet at $\sim 9000$ \AA; Fig.~\ref{fig:color-lightcurves}).  A perhaps more dramatic difference between AIC transients produced by a disk around the newly formed NS and other WD-related events are the large outflow speeds $\sim 0.1$c in AIC, which leads to highly velocity-broadened lines.  

\begin{figure}
\centering
\begin{overpic}[scale=0.30]{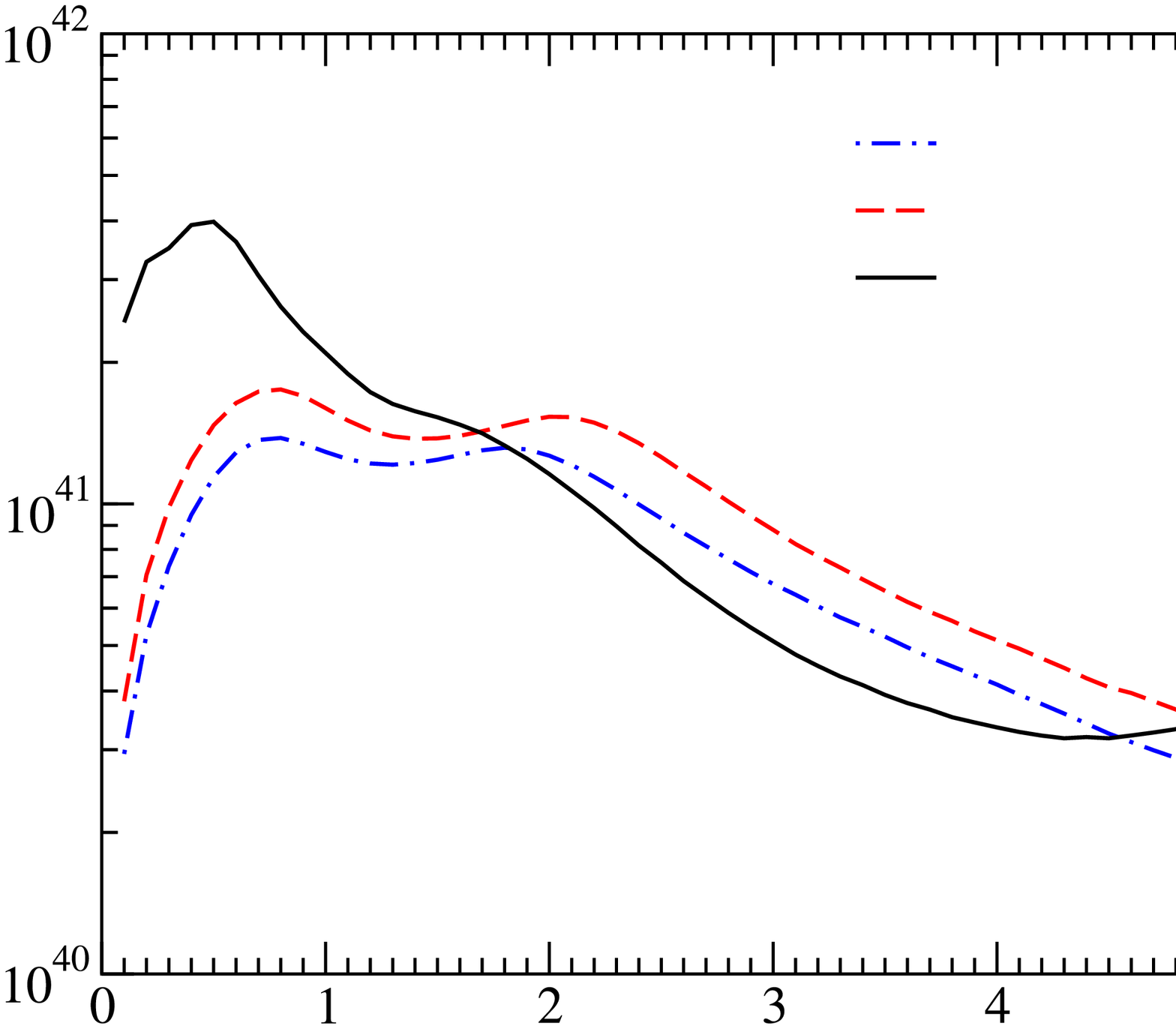}
\put(113,15){Time (d)}
\put(8,40){\begin{sideways}Bolometric Luminosity $\left(\textrm{ergs s}^{-1}\right)$\end{sideways}}
\put(153,149){$Y_e : 0.425 - 0.55$}
\put(153,139.5){$Y_e : 0.45 - 0.55$}
\put(153,130.5){$\epsilon_{\textrm{therm}} = 0.5$}
\put(48,80){AIC}
\put(48,146){NS-NS merger}
\end{overpic}
\caption{Comparison of the bolometric light curves of the transients from AIC (this paper) and NS mergers \citep{Metzger10}.  Both models were calculated for a total ejecta mass of $M_{\rm tot} = 10^{-2}M_{\sun}$.  The NS-NS merger transient is produced by the decay of a large number of neutron-rich nuclei created by r-process nucleosynthesis in the ejecta; \citet{Metzger10} estimated 
a thermalization efficiency of $\epsilon_{\textrm{therm}} \sim 0.5$ for these decays.  This figure highlights that distinguishing NS merger and AIC transients using lightcurve evolution alone will likely require early-time observations at $\lesssim$ 1 day.
\label{fig:aic-ns-bol-lightcurves}}
\end{figure}

The speed of the ejecta from NS mergers, as in AIC, is expected to be large.  A distinguishing feature in this case, however, is that NS merger ejecta is more neutron-rich than AIC ejecta, because in AIC the accretion disk material that ultimately becomes unbound is first irradiated by electron neutrinos from the proto-neutron star (\citealt{MPQ09-Neutron,MPQ09-Nickel}).  Although the current line data for the $r-$process nuclei produced in NS merger ejecta is insufficient to make concrete spectroscopic predictions, it is safe to say that NS mergers are unlikely to resemble any spectral class of SNe yet detected.  AIC and NS merger transients may also be distinguished by their light curve evolution.  In Figure \ref{fig:aic-ns-bol-lightcurves} we compare the bolometric light curves from AIC (this paper) and NS mergers (from \citealt{Metzger10}) assuming an ejecta mass $M_{\rm ej} = 10^{-2}M_{\sun}$ in both cases.  Although the overall brightness and late-time evolution is similar, the NS merger light curve shows an earlier, more pronounced peak.  This difference results from the different radioactive heating rate in the two cases.  AIC transients are powered by the decay of $^{56}$Ni, which produces an approximately constant heating rate prior to the $^{56}$Ni half-life $\sim 6$ days.  NS merger transients, by contrast, are powered by a large number of decaying nuclei with a wide distribution of half-lives, resulting in a total heating rate that decreases approximately as a power-law $\dot{E} \propto t^{-1.2-1.4}$.   NS merger transients peak earlier simply because the radioactive heating is concentrated at early times.  Distinguishing between NS mergers and AIC based on light curve evolution alone would thus probably require catching the event prior to maximum. 

Several sub-luminous, rapidly-evolving Type I supernovae have recently been discovered, which could in principle be related to AIC.  These include SN 2008ha (\citealt{Valenti+09}; \citealt{Foley+09}; \citealt{Foley+10}), 2005E (\citealt{Perets+09}), and 2002bj (\citealt{Poznanski+10}).  In all of these events, the low $^{56}$Ni masses inferred from their peak luminosities ($\sim 10^{-3}-0.1M_{\sun}$) are largely consistent with the quantity predicted from AIC.  However, the presence of low and intermediate mass elements (e.g. Ca, C, O, He) in the spectra of these events with relatively low velocities ($\sim 2000-10000$ km s$^{-1}$) rules out the standard Ni-rich AIC scenario outlined in this paper.  
  
As discussed by \citet{MPQ09-Nickel}, the timescale, velocity, and composition of the AIC transient can be modified if a substantial quantity of WD material remains at large radii following AIC, either because the WD is still collapsing or because some matter remains centrifugally supported in a remnant disk created by a super-M$_{\rm Ch}$ WD-WD merger (e.g. \citealt{Yoon+07}).  In the latter case, depending on the mass and composition of the WDs, up to $\sim 0.1M_{\sun}$ in C, O, Ne, or He could remain at a radius $\sim 10^{9}$ cm.  Once the SN explosion or Ni disk wind impacts the remnant material, it will (1) shock heat the material to a few times $10^{9}$ K, which may synthesize intermediate mass elements but will not disassociate the Ni and will leave some unburned WD material; (2) slow the ejecta from $v \sim 0.1$ c to a few thousand km s$^{-1}$.  A combination of slower ejecta and higher opacity would lengthen the duration of the light curve, making such an event easier to detect.  Such an ``enshrouded AIC'' scenario may therefore still be capable of explaining events such as SN 2008ha, 2005E, and 2002bj.  In order to make a quantitative comparison to observations, however, hydrodynamic simulations will be required to determine the dynamics and nucleosynthetic yield of the complex interaction between the prompt SN, the Ni-rich disk wind, and the remnant WD material.   Moreover, the timescale of AIC relative to the WD-WD merger itself is uncertain; any significant delay would likely lead to less debris at large radii and a ``cleaner" event analogous to that calculated in this paper.

On this note it is also important to distinguish between AIC due to accretion from a binary companion, as considered in this paper, and electron-capture supernova in AGB stars with O-Ne-Mg cores (for recent work, see \citealt{Kitaura+06}; \citealt{Wanajo+09}).  Although the collapse mechanism in these two cases is similar, an important difference in the AGB case is the presence of an extended hydrogen envelope.  This additional overlying material both increases the $^{56}$Ni yield of the explosion and extends the duration of the supernova due to its larger opacity (\citealt{Wanajo+09}).  Although the lack of hydrogen in the spectra of SNe 2008ha, 2005E, and 2002bj rules out electron capture SNe in these events, electron capture events may explain slower evolving sub-luminous SN such as 2008S (e.g.~\citealt{Prieto+08}; \citealt{Thompson+09}; \citealt{Pumo+09})

Our calculations could be improved in a number of ways.  In particular, future work will require more realistic multi-dimensional models in which the composition is calculated from a hydrodynamic simulation of the disk evolution beginning with free nucleons.   This is important for determining the final composition as a function of radius, which in turn can influence the spectroscopic predictions.  

\section*{Acknowledgments}  We thank Tony Piro for useful conversations.   SD and EQ are supported in part by the David \& Lucile Packard Foundation.  Support for EQ was also provided by the Miller Institute for Basic Research in Science, University of California  Berkeley.   Support for BDM was provided by NASA through Einstein Postdoctoral Fellowship grant number PF9-00065 awarded by the Chandra X-ray Center, which is operated by the Smithsonian Astrophysical Observatory for NASA under contract NAS8-03060.  Support for DK was provided by NASA through Hubble fellowship grant HST-HF-01208.01-A awarded by the Space Telescope Science Institute, which is operated by the Association of Universities for Research in Astronomy, Inc., for NASA, under contract NAS 5-26555. This research has also been supported in part by the DOE SciDAC Program (DE-FC02-06ER41438). Support for RT and PN was provided by the Director, Office of Science, Office of High Energy Physics of the U.S. Department of Energy under Contract No. DEAC02-05CH11231.

\bibliographystyle{mn2e}
\bibliography{bibfile}

\label{lastpage}

\end{document}